\documentclass[aps,prl,twocolumn,showpacs]{revtex4}

\usepackage{mathptm}    
\usepackage{dcolumn}                    
\usepackage{bm}                         
\usepackage{graphicx}

\usepackage{times}

\begin{document}

\title{Resonant impurity scattering in the $\pm$s-wave state of the Fe-based superconductors}

\author{Yunkyu Bang}
\email[To whom the correspondences should be addressed:
]{ykbang@chonnam.ac.kr} \affiliation{Department of Physics,
Chonnam National University, Kwangju 500-757, and Asia Pacific
Center for Theoretical Physics, Pohang 790-784, Korea}

\author{Han-Yong Choi}
\affiliation{Department of Physics and Institute for Basic Science
Research, SungKyunKwan University, Suwon 440-746, Korea}

\begin{abstract}
We study the impurity scattering on the $\pm$s-wave
superconductor, with realistic parameters for the Fe-pnictide
superconductors. Using $\mathcal{T}$-matrix method, generalized
for the two bands, we found that impurity scattering of the
unitary limit forms off-centered bound states inside of the
superconducting gap, which modifies, surprisingly, the density of
states (DOS) of a fully opened gap to a V-shaped one as in the
case of a d-wave superconductor. This behavior provides coherent
explanations to the several conflicting experimental issues of the
Fe-pnictide superconductors: the V-shaped DOS but with an
isotropic gap observed in the photoemission and tunneling
experiments; the power law behavior of the nuclear spin-lattice
relaxation rate ($1/T_1 \approx T^{\alpha}$ ; $\alpha \approx 3$),
down to very low temperatures.

\end{abstract}


\date{\today}
\maketitle

{\it Introduction -} The recent discovery of Fe-based
superconducting (SC) compounds \cite{Kamihara,Chen}, has greatly
spurred the research activity of the superconductivity. With a
discovery of new superconducting material, the most impelling
question is to determine the SC gap symmetry. Identifying the
pairing symmetry, the search for the possible pairing mechanisms
can be more easily advanced. Various SC properties were already
measured to determine the gap symmetry of these materials but
there exist serious conflicts among data. Just list some of them,
tunneling spectroscopy of Ref.\cite{tunneling-d}, photoemission
measurement of Ref.\cite{photoemission-d}, and nuclear-
spin-lattice relaxation rate measurements \cite{T1, T1-Zheng} seem
to indicate a d-wave type gap with lines of node. On the other
hand, tunneling spectroscopy of Ref.\cite{tunneling-s},
photoemission measurement of Ref.\cite{photoemission-s}, specific
heat \cite{C(T)}, and the penetration depth measurements by
several groups \cite{pene} all support a fully opened s-wave type
gap.

On the theoretical side, it is almost agreed on that the pairing
mechanism is non-phononic\cite{phonon}, and most probably of a
magnetic origin \cite{Dai,Bao}. This conclusion is consistent with
the overall phase diagram obtained by neutron scattering
\cite{Dai}, where the SC phase starts to develop when the
antiferromagentic (AFM) order disappears, suggesting an universal
mechanism for unconventional superconductivity of the compounds
with d- and f-electron elements. For the Fe-based SC compounds, an
AFM correlation induced interaction with the specific band
structure - in particular, two topologically distinct hole and
electron bands widely separated  in the Brilluion Zone - favors to
develop so-called $\pm$s-wave SC state, first proposed by Mazin et
al. \cite{Mazin}, and reconfirmed theoretically by several authors
\cite{Kuroki,Eremin,Bang-Fe,DHLee}.

Then the key question is : Is the $\pm$s-wave state consistent
with all experiments ? Up to now, the $\pm$s-wave state is the
most natural candidate to understand the penetration depth
behavior \cite{pene}. However, the nuclear-spin-lattice relaxation
rate $1/T_1$ experiments by several groups \cite{T1,T1-Zheng} all
support a nodal gap SC state. Recently, several groups
\cite{recent,Parish}, including the present authors
\cite{Bang-Fe}, proposed that the unusual interband coherence
factor, unique to the $\pm$s-gap state due to the opposite signs
of the SC order parameter (OP) between the bands \cite{S-pi}, can
explain the absence of Hebel-Slichter peak  as well as an
approximate power law behavior of $1/T_1$, with a help of
impurities.
While this is an important observation, the calculations of these
works \cite{recent,Bang-Fe,Parish} are not completely satisfactory
to reproduce the $T^3$ power law \cite{T3note}. Also the works of
Ref.\cite{recent} considered only the interband scattering process
while the works of Ref.\cite{Bang-Fe,Parish} showed that both the
interband and intraband processes have comparable contributions
and hence both processes should be treated on equal footing.
Therefore, it needs more quantitative and systematic studies to
settle the issue; in particular, how robust and how low
temperatures the power law behavior can extend to, and the nature
of impurities.

In this paper, we employed the $\mathcal{T}$-matrix approximation,
generalized to the $\pm$s-wave state of the two band model, to
study the effects of impurities from weak (Born) to strong
(unitary) scattering limit. The $\mathcal{T}$-matrix approximation
has been successfully applied to the various unconventional
superconductors such as heavy fermion \cite{T-hf} and
high-temperature superconductors \cite{T-htc}. For example, it
predicts a resonant bound state by unitary impurity scatterer
inside the d-wave SC gap, which was crucial to explain the
penetration depth of HTSC \cite{T-htc}, $1/T_1$ experiments of
Pu-115 superconductor\cite{Pu}, etc. The key principle of forming
a resonance bound sate in the d-wave gap is that the sign-changing
d-wave SC OP guarantees the absence of the renormalization of the
anomalous selfenergy due to impurity scattering. Therefore, we
expect a similar mechanism to work with the sign-changing
$\pm$s-wave superconductors. However, there are important
differences: (1) the cancellation of the anomalous selfenergy
would not be perfect unless the sizes of s$+$ and s$-$ gap and
their corresponding DOSs N$_h$(0) and N$_e$(0) are exactly equal;
(2) the DOS of the pure state is not linearly vanishing as in the
d-wave case but a fully gapped one. Considering these differences,
it requires a quantitative and transparent investigation to
understand the effects of impurity scattering on the $\pm$s-wave
state.

{\it Formalism -} Recently we had proposed a minimal two band
model with a phenomenological magnetic interaction for the
Fe-based superconductors \cite{Bang-Fe}. Assuming two SC order
parameters, $\Delta_h$ and $\Delta_e$ on each band, the two
coupled gap equations are written as \vspace{-0.3cm}

\begin{eqnarray}
\Delta_h (k)  &=&  \\ \nonumber  - & \sum_{k^{'} } & [V_{hh}
(k,k^{'})   \chi_h (k^{'}) + V_{he} (k,k^{'}) \chi_e (k^{'})], \\
\nonumber
\Delta_e (k)  &=&  \\ \nonumber  - & \sum_{k^{'} }& [V_{eh}
(k,k^{'})   \chi_h (k^{'}) + V_{ee}
(k,k^{'})  \chi_e (k^{'})]. \\
\end{eqnarray}

\noindent where $V_{h,e} (k,k^{'})$ is the phenomenological
pairing interaction originating from the AFM correlation. The
above gap equations indeed provide the $\pm$s-wave state
($\Delta_h$ and $\Delta_e$ have the opposite signs) as the best
solution \cite{Bang-Fe}.
The impurity effects enter through the pair susceptibility in the
following way.

\begin{eqnarray}
\chi_{h,e}(k) &=& T \sum _n N(0)_{h,e} \int _{-\omega_{AFM}}
^{\omega_{AFM}} d \xi \frac{ \tilde{\Delta}_{h,e}(k) } {
\tilde{\omega}_n^2 +\xi^2 + \tilde{\Delta}_{h,e} ^2 (k)}
\end{eqnarray}

\noindent where $N(0)_{h,e} $ are the DOS of the hole and electron
bands, respectively, and $\omega_{\rm AFM}$ is the cutoff energy
of the pairing potential $V(q)$. $\tilde{\omega}_n =\omega_n +
\Sigma^0 _h(\omega_n) + \Sigma^0 _e(\omega_n)$ and
$\tilde{\Delta}_{h,e} = \Delta_{h,e} + \Sigma^1 _{h} (\omega_n) +
\Sigma^1 _{e} (\omega_n)$ (where $\omega_n= \pi T  (2n +1)$), and
the impurity induced selfenergies are calculated with
$\mathcal{T}$-matrices as $\Sigma_{h,e} ^{0,1} (\omega_n) = \Gamma
\cdot \mathcal{T}^{0,1} _{h,e} (\omega_n)$; $\Gamma=n_{imp}/\pi
N_{tot}$ where $n_{imp}$ is the impurity concentration and
$N_{tot}=N_h(0) +N_e(0)$ is the total DOS. The
$\mathcal{T}$-matrices $\mathcal{T}^{0,1}$ are the Pauli matrices
$\tau^{0,1}$ components in Nambu space. We can set $\mathcal{T}^3
=0$ assuming particle-hole symmetry and $\mathcal{T}^2=0$ due to
U(1) symmetry, respectively, without the loss of generality.
Definitions of $\mathcal{T}^{0,1} _{h,e} (\omega_n)$ are standard
\cite{T-hf} in the literatures, but now need a generalization for
the two bands as follows.

\begin{eqnarray}
\mathcal{T}^{i} _{a} (\omega_n) &=& \frac{G^{i} _{a} (\omega_n)}{D} ~~~~~(i=0,1; ~~a=h,e), \\
D &=& c^2 +[G^0 _h + G^0 _e]^2 + [G^1 _h + G^1 _e]^2,\\
G^0 _a (\omega_n) &=& \frac{N_a}{N_{tot}} \left\langle
\frac{\tilde{\omega}_n}
{\sqrt{\tilde{\omega}_n^2 + \tilde{\Delta}_{a} ^2 (k) }} \right\rangle,\\
G^1 _a (\omega_n) &=& \frac{N_a}{N_{tot}} \left\langle
\frac{\tilde{\Delta}_{a}} {\sqrt{\tilde{\omega}_n^2 +
\tilde{\Delta}_{a} ^2 (k) }},  \right\rangle
\end{eqnarray}

\noindent where $c=\cot \delta_0$ is a convenient measure of
scattering strength, with c=0 in the unitary limit and $c > 1$ in
the Born limit scattering. $\langle ...\rangle$ denotes the Fermi
surface average. The above four $\mathcal{T}$-matrices,
$\mathcal{T}^{i} _{a}$ are numerically solved together with the
coupled gap equations Eq.(1) and Eq.(2) in Matsubara frequencies.
Then Pade-approximant method\cite{Vidberg} is used to obtain the
real and imaginary part of selfenergies $\Sigma_{h,e} ^{0,2}
(\omega + i \eta)$, with which various physical quantities can be
straightforwardly calculated.

Note Eq.(5), which is the denominator of $\mathcal{T}$-matrices.
The last term of it, $[G^1 _h + G^1 _e]$, would exactly vanish for
a d-wave and that is the technical reason for the formation of the
resonance bound state at zero energy when $c=0$, the unitary limit
scattering. For the $\pm$s-wave case, $G^1 _h$ and  $G^1 _e$ have
opposite signs, therefore a large cancellation in $[G^1 _h + G^1
_e]$ occurs but never be perfect unless $\Delta_e = - \Delta_h$
and $N_h (0) = N_e (0)$. With an incomplete cancellation, the
finite remnant acts as weakening the scattering strength $c$ (it
means increasing the effective value of $c$). For the middle term
$[G^0 _h + G^0 _e]$, $G^0 _h$ and $G^0 _e$ are always the same
sign, so that the normal scattering process is additive with
number of bands.

With the typical band structure of the Fe-based
pnictides\cite{band}, $N_h (0)$ and $ N_e (0)$ are not equal. Then
the coupled gap equations Eq.(1) and Eq.(2), using realistic band
parameters, resulted in the gap solutions such that  $|\Delta_e| /
|\Delta_h| \approx 2.5$  with $N_h (0)/N_e (0) \approx 2.6$ in the
previous calculations \cite{Bang-Fe}.
Notice the substantial difference of the sizes of the gap for the
hole and electron bands and the inverse relation between two
ratios $N_h (0)/N_e (0)$ and $|\Delta_h| / |\Delta_e|$. This
inverse relation between the gap sizes and the DOS sizes -- i.e,
$|\Delta_h| < |\Delta_e|$ for $N_h (0) > N_e (0)$ and vice versa
-- is a generic property of the interband pairing model
\cite{note1}.
For most calculations in this paper, we used the above parameters.

In this paper, we are particularly interested in the effects of
impurity scattering on the nuclear spin-lattice relaxation rate
$1/T_1$ of the $\pm$s-wave state, which is calculated by

\begin{eqnarray}
\label{1 o T1} \frac{1}{T_1} &\propto&  - T \int_0 ^{\infty} d
\omega \frac{\partial f_{FD} (\omega)}{\partial \omega} \Biggl
\lbrace \sum_{a=h,e}
 N_a ^2(0) \Bigg[  \left\langle Re
   \frac{\tilde{\omega}}{\sqrt{\tilde{\omega}^2-\tilde{\Delta}_a ^2(k)}}
  \right\rangle_{k}^2
  \nonumber \\
  &+& \left\langle Re \frac{\tilde{\Delta}_a(k)}{\sqrt{\tilde{\omega}^2-\tilde{\Delta}_a
^2(k)}} \right\rangle_{k}^2 \Bigg]  + ~~ 2 N_h (0) N_e (0) \Bigg[
  \nonumber \\
   & &  \left\langle Re
     \frac{\tilde{\omega}}{\sqrt{\tilde{\omega}^2-\tilde{\Delta}_h ^2(k)}}
    \right\rangle_{k} \left\langle Re
     \frac{\tilde{\omega}}{\sqrt{\tilde{\omega}^2-\tilde{\Delta}_e ^2(k^{'})}}
    \right\rangle_{k^{'}}
 \nonumber \\    &+& \left\langle Re
     \frac{\tilde{\Delta}_h(k)}{\sqrt{\tilde{\omega}^2-\tilde{\Delta}_h ^2(k)}}
    \right\rangle_{k} \left\langle Re
     \frac{\tilde{\Delta}_e(k^{'})}{\sqrt{\tilde{\omega}^2-\tilde{\Delta}_e ^2(k^{'})}}
    \right\rangle_{k^{'}} \Bigg] \Biggr \rbrace.
\end{eqnarray}

\noindent where $\tilde{\omega}$ and
$\tilde{\Delta}_{h,e}(\omega)$ are the real frequency quantities
obtained from $\tilde{\omega}_n$ and
$\tilde{\Delta}_{h,e}(\omega_n)$ by analytic continuation
$\omega_n \rightarrow \omega + i \eta$. The above formula of
$1/T_1$ contains three scattering channels: two intraband
scattering channels from the hole band and electron band,
respectively, and one interband scattering channel between the
hole and electron bands. The unique feature of the $\pm$s-wave
state is the interband scattering channel. Having the opposite
signs for $\Delta_h$ and $\Delta_e$, this term (the last term in
Eq.(\ref{1 o T1})) substantially cuts out the relaxation rate
below T$_c$ as shown by recent papers
\cite{recent,Bang-Fe,Parish}. However, we emphasized that this
interband term is not sufficient enough to completely wash out the
coherence peak, in particular, when the other two intraband
scattering processes are correctly included \cite{Bang-Fe}.
For the temperature dependence of the gaps $\Delta_{h,e}(k,T)$, we
use a phenomenological formula,
$\Delta_{h,e}(k,T)=\Delta_{h,e}(k,T=0) \tanh (\beta
\sqrt{T_{c}/T-1})$. By choosing the values of $\Delta _{h,e}
^{max} / T_c$, we can partially take into account of the effect of
the strong coupling superconductivity. We fix $\beta=1.74$ as
usual, which is not a sensitive parameter for final results.

\begin{figure}
\noindent
\includegraphics[width=70mm]{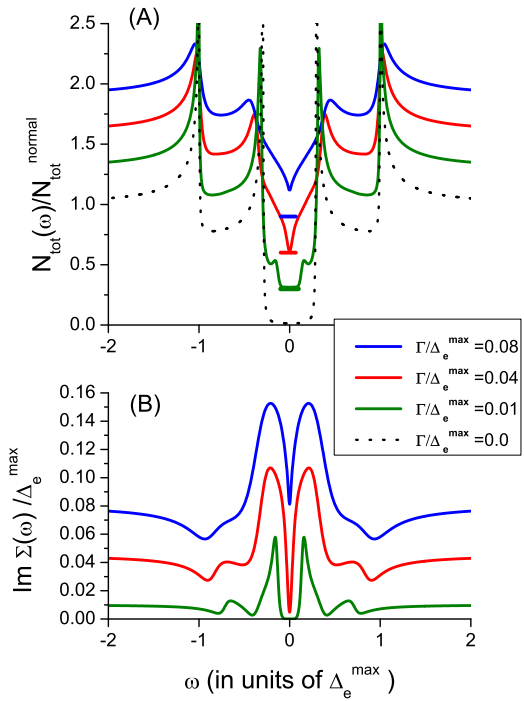}
\caption{(Color online) (a) Normalized DOS $N_{tot}(\omega)$ for
different impurity concentrations, $\Gamma / \Delta_e= 0.0, 0.01,
0.04, 0.08$. Thin dotted line is of the pure state for comparison
and other lines are offset for clarity (the zero baselines of the
offset are marked by the narrow horizontal bars of the
corresponding colors). (b) Impurity induced selfenergies $Im
\Sigma^0 _{tot} (\omega) = Im \Sigma^0 _h + Im \Sigma^0 _e$ with
the same parameters as in (a). These curves are not offset.
\label{fig1}}
\end{figure}

{\it Results - } All energy scales are normalized by $|\Delta_e|$
in this paper. Fig.1 (a) shows the total DOS of two bands with
different impurity concentrations $\Gamma / \Delta_e= 0.0, 0.01,
0.04, 0.08$ of the unitary scatterer (c=0), and Fig.1(b) shows the
corresponding impurity induced selfenergy $Im \Sigma^0 _{tot}
(\omega) = Im \Sigma^0 _h + Im \Sigma^0 _e$. Fig.1 (a) shows how
the fully opened gap of the pure state is filled with impurity
states; the pattern of filling is very unusual and the $\Gamma /
\Delta_e= 0.04$ case displays a perfect V-shape DOS down to zero
energy as in a d-wave SC gap. The origin of this behavior is
easily seen in Fig.1(b); the impurity bound state is never formed
at zero energy but away from it (even in the unitary limit)
because of the incomplete cancellation of $[G^1 _h + G^1 _e]$, so
{\it the full gap around $\omega=0$ is protected} until this
off-centered impurity band spills over to the zero energy with
increasing the impurity concentration. When it touches the zero
energy limit, the superconductor behaves gapless as in a pure
d-wave superconductor, and this happens with the critical impurity
concentration $\Gamma_{crit}$ ($=0.04 \Delta_e$ for our specific
model parameters). Increasing the impurity concentration beyond
$\Gamma_{crit}$, the DOS still keeps the V-shape but now $N_{tot}
(\omega=0)$ obtains a finite value (see the blue curve of
$\Gamma=0.08 \Delta_e$ case in Fig.1(a)).

This manner of evolution of the DOS with the impurity
concentration results in the following consequences: (1) Beyond
the critical impurity concentrations, direct measurements of the
DOS at low temperature such as photoemission and tunneling
spectroscopy would see a V-shape DOS, but at the same time would
be extracting an isotropic gap \cite{tunneling-d,photoemission-s};
(2) Temperature dependence measurement such as $1/T_1 (T)$ would
see three different types of behavior. First, when
$\Gamma=\Gamma_{crit}$ ($\Gamma=0.04 \Delta_e$ case in Fig.1(a)),
the system sees the linear in $\omega$ DOS for whole temperature
region of 0 $<$ T $<$ T$_c$. Second, when $\Gamma
>\Gamma_{crit}$ ($\Gamma=0.08 \Delta_e$ case in
Fig.1(a)), the linear in $\omega$ DOS will prevail in the high
temperature region, but at low temperatures the finite DOS of
$N_{tot} (\omega=0)$ makes the system a gapless superconductor.
Finally, when $\Gamma < \Gamma_{crit}$ ($\Gamma=0.01 \Delta_e$
case in Fig.1(a)), the system always behaves as a fully opened gap
superconductor although the gap is weakened by impurities. This
variation of DOS with the impurity concentration will be reflected
in the behavior of $1/T_1 (T)$ as will be shown below.
In passing, note that $Im \Sigma^0 _{tot} (\omega)$ in Fig.1(b)
shows two peaks on each side of the $\omega$-axis. Apparently, a
smaller energy peak ($\omega \sim 0.2 \Delta_e$) -- but with a
larger spectral density due to the larger DOS of the hole band --
is formed inside of the small gap $\Delta_h$ and the larger energy
peak ($\omega \sim 0.7 \Delta_e$) is formed inside of the larger
gap $\Delta_e$.

\begin{figure}
\noindent
\includegraphics[width=70mm]{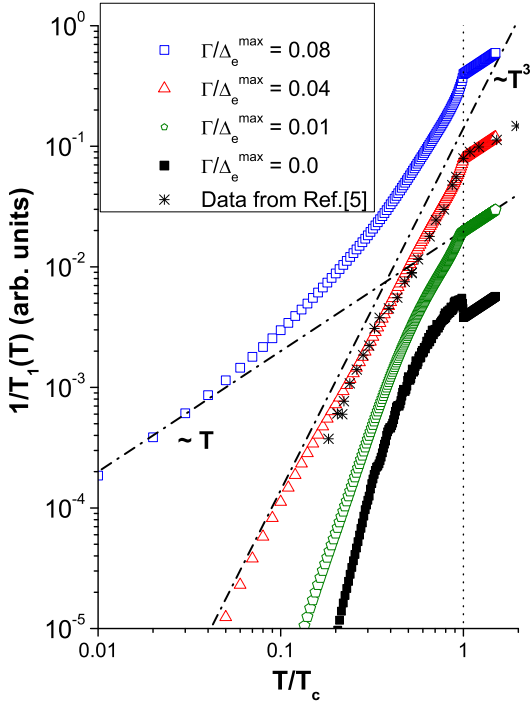}
\caption{(Color online) Calculated $1/T_1 (T)$ for different
impurity concentrations, $\Gamma / \Delta_e= 0.0, 0.01, 0.04,
0.08$ and with $2 \Delta_h / T_c$=3.0. Experimental data is from
Ref.\cite{T1-Zheng}. The curves are offset for clarity.
\label{fig2}}
\end{figure}

Figure 2 shows the calculations of $1/T_1(T)$ with the variation
of the impurity concentration using the same parameters as in
Fig.1. It is clear that the puzzling $T^3$ behavior of $1/T_1$ can
be understood with the $\pm$s-wave; it has the same origin as in
the d-wave gap, i.e., the linearly rising DOS.
With $\Gamma = \Gamma_{crit}=0.04 \Delta_e$, the $T^3$ behavior
extends to the lowest possible temperatures as expected. With
$\Gamma > \Gamma_{crit}$, the $T^3$ behavior occurs only at high
temperatures and at lower temperatures the system probes the
finite DOS of $N_{tot} (\omega=0)$, hence displaying the
$T$-linear behavior of $1/T_1$. With $\Gamma < \Gamma_{crit}$, the
system should display a full gap behavior below $T_c$, but
somewhat weakened by impurities. As a consequence, $1/T_1$ shows,
in this case, a much weakened exponential drop for the extended
temperature region below $T_c$.
This wide range of variation occurs with the impurity
concentration $0 < \Gamma/\Delta_e < 0.08$ and the reduction of
$T_c$ due to impurities is less than $10 \%$; $\delta T_c / T_c
^0$ is proportional to $(\Gamma/\Delta_e)/[c^2+1]$.
Finally, we emphasize that in order to capture this systematic
evolution of $1/T_1$ with impurity concentration, it is absolutely
necessary to include both interband and intraband scattering
process on equal footing.

Figure 3 shows an artificial case of the equal size $\pm$s-wave
gap ($|\Delta_e|= |\Delta_h|$ and $N_h (0)=N_e (0)$). Because of
the perfect cancellation of $[G^1 _h + G^1 _e]$ term in the
denominator of the $\mathcal{T}$-matrices, any small amount of the
impurity concentration immediately induces the bound state at zero
energy as in the case of a d-wave superconductor. But this bound
state spectral density is isolated inside of a full gap (see
Fig3.(b)). Therefore, the superconductivity still remains as a
full gap superconductor, until the bound state spectral density
grows and touches the edge of the gap with increasing the impurity
concentration. The inset of Fig.3(b) shows the imaginary part of
the impurity selfenergy $Im \Sigma_{tot} (\omega)$, clearly
showing the bound state centered at $\omega=0$, which should be
contrasted with the off-centered bound state for the unequal size
gap case (see Fig.1(B)).

As a result, the SC state in this case shows an activated behavior
for the high temperature region below T$_c$ and the impurity
induced DOS only starts being sensed at very low temperatures.
This behavior is well captured with $1/T_1 (T)$ (Fig.3(A)).
It shows that even the Hebel-Slichter peak around $T_c$ is not
completely suppressed despite the perfect cancellation of $[G^1 _h
+ G^1 _e]$, because at or near $T_c$ the impurity effect is only
weakly sensed by the system. This result emphasizes that the
assumption of the equal size gaps, as done in Ref.\cite{recent},
results in a qualitatively different physics and it appears not
consistent with experiments \cite{T1,T1-Zheng}. For comparison, we
plot the experimental data by Kawasaki et al.\cite{T1-Zheng}
together with the theoretical result in Fig.3(a).

\begin{figure}
\noindent \includegraphics[width=90mm]{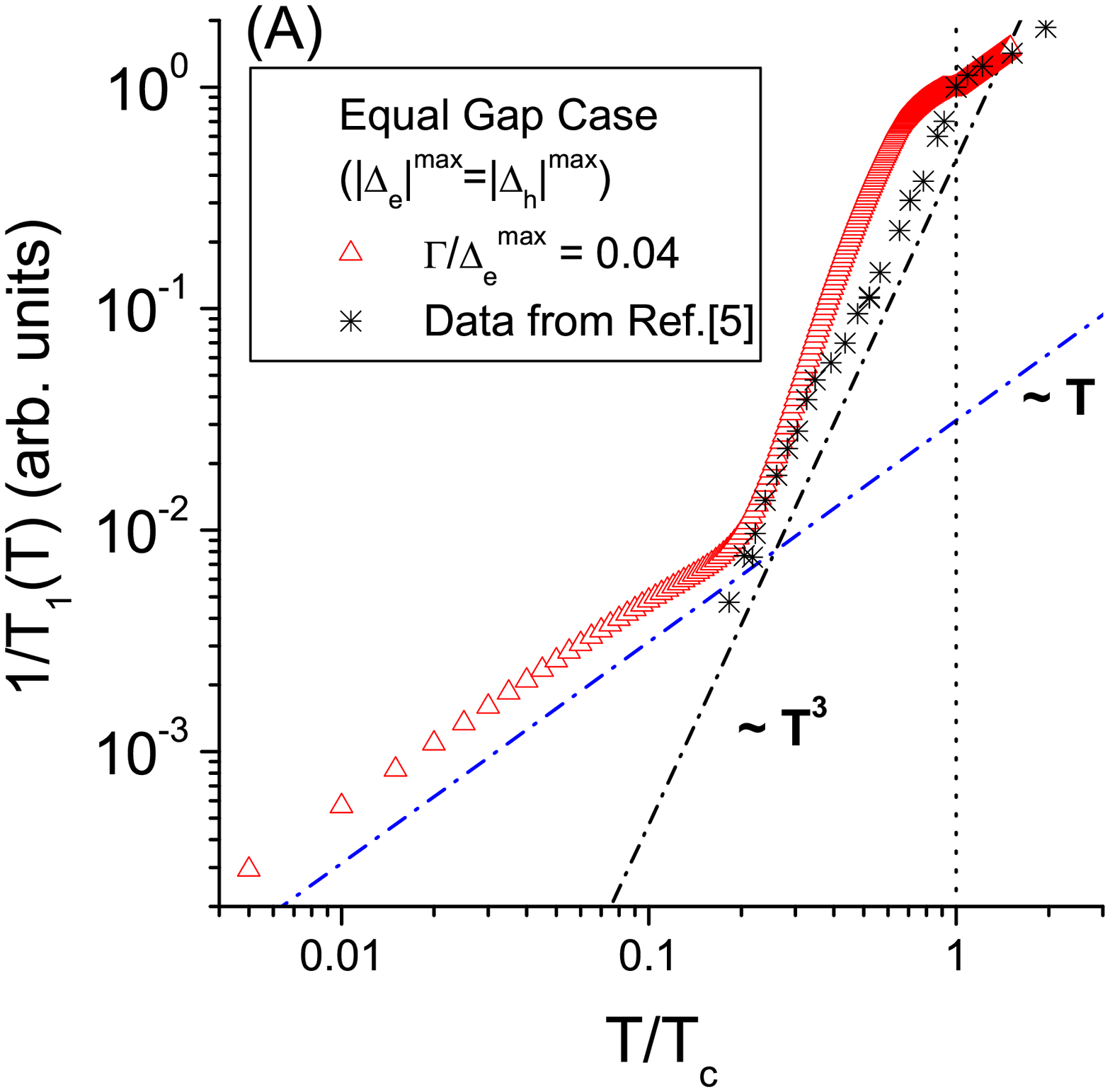}
\includegraphics[width=90mm]{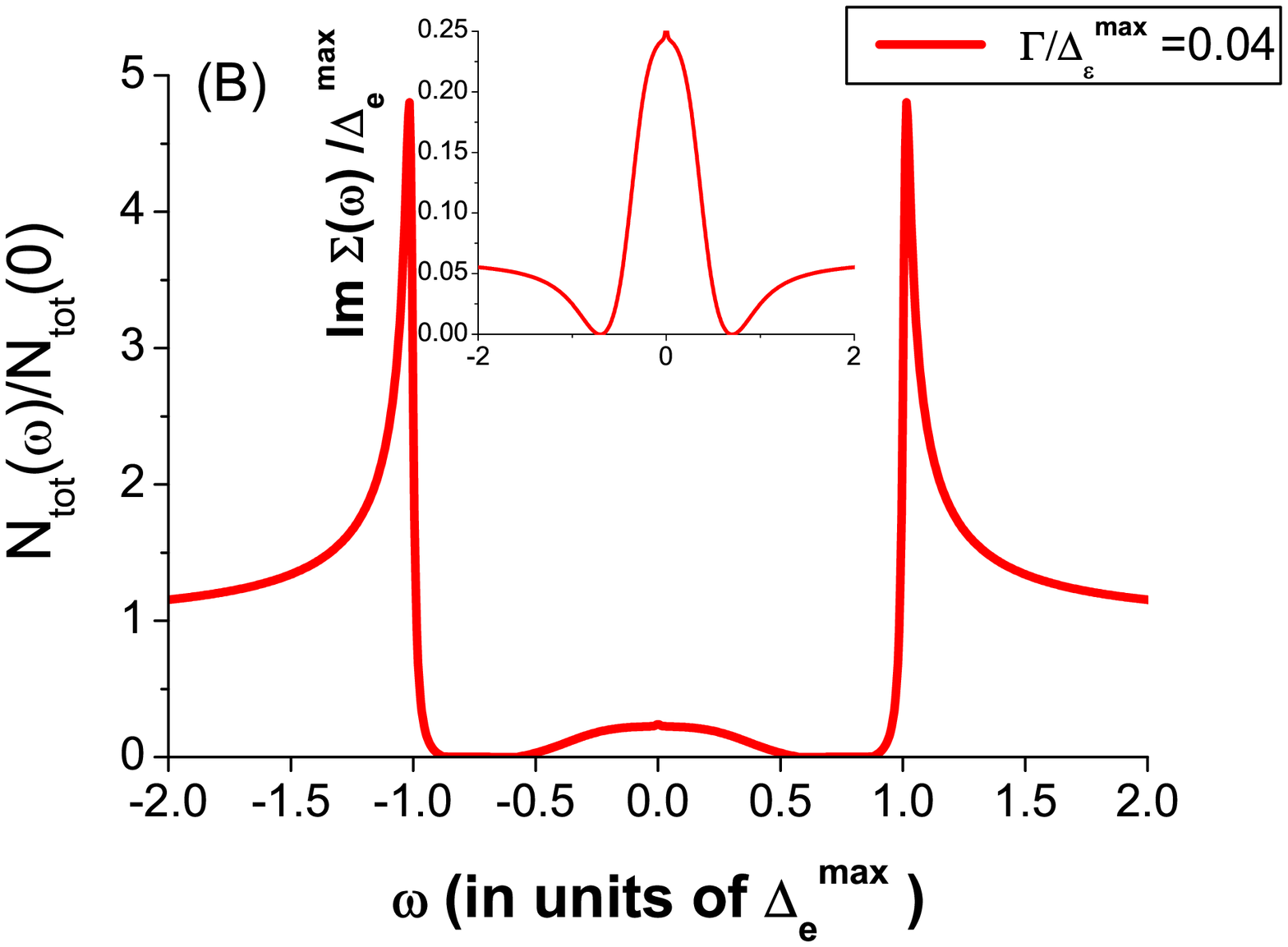} \caption{(Color online)
The case of the equal size $\pm$s-wave gap, ($\Delta_e= -\Delta_h$
and $N_h (0)=N_e (0)$). (a) $1/T_1 (T)$ with $2 \Delta_h
/T_c$=3.0. Experimental data is from Ref.\cite{T1-Zheng}. (b)
Normalized DOS N$_{tot}(\omega)$. Inset: Impurity induced
selfenergy $Im \Sigma^0 _{tot} (\omega)$. \label{fig3}}
\end{figure}

{\it Conclusion -} In conclusion, we studied the effect of
impurities, in particular, of the strong scattering limit, on the
$\pm$s-wave superconductor with a generalized $\mathcal{T}$-matrix
method. The unique and generic feature of the $\pm$s-wave
superconductor, i.e., {\it the opposite signs of the gaps but with
unequal sizes}, results in the off-centered impurity bound states
inside the gaps. With the variation of the impurity concentration,
the DOS, N$_{tot}(\omega)$, evolves systematically from a fully
gapped one to a V-shape one. We showed that several conflicting
experimental issues such as photoemission, tunneling spectroscopy,
and $1/T_1$ are coherently explained with the $\pm$s-wave SC state
with the resonant impurity scattering.

{\it Acknowledgement - } This work was supported by the KOSEF
through the Grants No. KRF-2007-521-C00081 (YB), No.
KRF-2007-070-C00044 (YB,HYC), and Basic Research Program Grant No.
R01-2006-000-11248-0 (HYC). We thank Guo-qing Zheng for useful
discussions and sending us their experimental data.


\begin{references}

\bibitem{Kamihara}
Y. Kamihara et al., J. Am. Chem. Soc., {\bf 128}, 10012 (2006); Y.
Kamihara et al., J. Am. Chem. Soc., {\bf 130}, 3296 (2008).

\bibitem{Chen}
G. F. Chen et al., Phys. Rev. Lett. {\bf 100}, 247002 (2008); G.
F. Chen et al., Nature {\bf 453}, 761 (2008).


\bibitem{tunneling-d}  Y. Wang et al., arXiv:0806.1986 (unpublished);
L. Shan et al., Europhys. Letters, {\bf 83}, 57004 (2008).

\bibitem{photoemission-d}
T. Sato et al., J. Phys. Soc. Jpn. {\bf 77}  063708 (2008).

\bibitem{T1-Zheng}
S. Kawasaki et al., Phys. Rev. B {\bf 78}, 220506(R) (2008).

\bibitem{T1}
K. Matano et al., Europhys. Lett. {\bf 83}  57001 (2008);  H.-J.
Grafe et al., Phys. Rev. Lett. {\bf 101}, 047003 (2008);  H.
Mukuda et al., J. Phys. Soc. Jpn. {\bf 77} (2008) 093704; Y. Nakai
et al., J. Phys. Soc. Jpn. {\bf 77} (2008) 073701.

\bibitem{tunneling-s}
T. Y. Chen et al., Nature (London), {\bf 453}, 1224 (2008).

\bibitem{photoemission-s}
H. Ding et al., Europhys. Lett. {\bf 83}, 47001 (2008); T. Kondo
et al., Phys. Rev. Lett. {\bf 101}, 147003 (2008);  L. Wray et
al., arXiv:0808.2185 (unpublished).

\bibitem{C(T)}
G. Mu et al., Chin. Phys. Lett., {\bf 25}, 2221 (2008).

\bibitem{pene}
L. Malone et al.,  arXiv:0806.3908 (unpublished);  K. Hashimoto et
al., Phys. Rev. Lett. {\bf 102}, 017002 (2009); C. Martin et al.,
arXiv:0807.0876 (unpublished).


\bibitem{phonon}
L. Boeri, O. V. Dolgov, A. A. Golubov, Phys. Rev. Lett. {\bf 101},
026403 (2008).

\bibitem{Dai}
C. de la Cruz et L., Nature (London) {\bf 453}, 899 (2008); J.
Zhao ET AL., Nature Materials {\bf 7}, 953 (2008).

\bibitem{Bao} Y.Qiu et al., Phys. Rev. Lett. {\bf 101}, 257002 (2008).

\bibitem{Mazin}
I.I. Mazin, D.J. Singh, M.D. Johannes, M.H. Du, Phys. Rev. Lett.
{\bf 101}, 057003 (2008).

\bibitem{Kuroki}
K. Kuroki et al., Phys. Rev. Lett. {\bf 101}, 087004 (2008).

\bibitem{Eremin}
M.M. Korshunov and I. Eremin, Phys. Rev. B {\bf 78}, 140509(R)
(2008).

\bibitem{Bang-Fe}
Y. Bang and H.-Y. Choi, Phys. Rev. B, {\bf 78}, 134523 (2008).

\bibitem{DHLee}
F. Wang, H. Zhai, Y. Ran, A. Vishwanath, Dung-Hai Lee, Phys. Rev.
Lett. {\bf 102}, 047005 (2009).


\bibitem{recent}
D. Parker, O.V. Dolgov, M.M. Korshunov, A.A. Golubov, I.I. Mazin ,
Phys. Rev. B {\bf 78}, 134524 (2008); A.V. Chubukov, D.V. Efremov,
I. Eremin, Phys. Rev. B {\bf 78}, 134512 (2008) .

\bibitem{Parish}
M. M. Parish, J. Hu, B. A. Bernevig, Phys. Rev. B {\bf 78}, 144514
(2008).

\bibitem{S-pi}
It is also possible to design a tunneling experiment to directly
probe the $\pi$ phase difference between the $\pm$s-gaps as
proposed in H.-Y. Choi and Y. Bang, arXiv:0807.4604 (unpublished).

\bibitem{T3note}
This $T^3$ power law behavior extracted from experimental data of
the Fe pnictide superconductors might be only an approximate
definition unless its origin is due to the gap symmetry as in the
d-wave SC. Our study shows, however, that it also arises from the
V-shape DOS as in the d-wave case; therefore the $T^3$ power law
of $1/T_1$ in the Fe pnictide superconductors has a rather firm
ground.

\bibitem{T-hf}
P. J. Hirschfeld, P. Wolfle, and D. Einzel, Phys. Rev. B {\bf 37},
83 (1988); A. V. Balatsky, I. Vekhter, and J.-X. Zhu, Rev. Mod.
Phys. {\bf 78}, 373 (2006) and see more references therein.

\bibitem{T-htc}
L. S. Borkowski and P. J. Hirschfeld Phys. Rev. B {\bf 49}, 15404
(1994).

\bibitem{Pu}
N. Curro, T. Caldwell, E.D.Bauer, L.A. Morales, M.J. Graf, Yunkyu
Bang, A.V. Balatsky, J.D. Thompson, J.L.Sarrao, Nature {\bf 434},
622 (2005).

\bibitem{Vidberg}
H.J Vidberg and J.W. Serene, J. of Low Temp. Phys. {\bf 29}, 179
(1977).

\bibitem{band}
D.J. Singh and M.-H. Du, Phys. Rev. Lett. {\bf 100}, 237003 (2008)
; C. Cao, P. J. Hirschfeld, H.P. Cheng, Phys. Rev. B {\bf 77},
220506(R) (2008); E. Manousakis, Jun Ren, E. Kaxiras,
arXiv:0806.3432 (unpublished).

\bibitem{note1}
Although for most of cases, the approximate relation  $N_h (0)
|\Delta_h| \approx N_e (0) |\Delta_e|$ holds, this is not a
rigorous identity. For more detail discussion, see
Ref.\cite{Bang-Fe}.



\end{references}
\end{document}